\documentclass{article}
\usepackage{frascatiphys}
\begin{document}
%
\title{
CP Violation Results in Charm
}
\author{
David Asner - for the CLEO Collaboration \\
{\em University of Pittsburgh, Department of Physics and Astronomy,}\\ 
{\em 3951 O'Hara St, Pittsburgh PA, 15260, USA} \\
}
\maketitle
\baselineskip=11.6pt
\begin{abstract}
Searches for CP violation in the charm sector from the E791, FOCUS, CLEO, BABAR and BELLE experiments are presented. Most analyses consider CP violation in two-body or quasi-two-body decays.
Preliminary results from CLEO and FOCUS using Dalitz-plot analyses are also presented.
\end{abstract}
\baselineskip=14pt
\section{Introduction}
The violation of charge-parity (CP) in charm decay requires two amplitudes
with different strong and weak phases that interfere to produce CP violating effects.
There are three distinct types of CP violation. (1) CP violation from a non-vanishing 
relative phase between the mass and width components of the mixing matrix 
usually called ``indirect''; (2) Direct CP violation due to
the two decay amplitudes having different weak phases; (3) Interference between decays
with and without mixing. The CP conserving phase shift is usually generated by QCD final state
interactions (FSI). In the Standard Model,
the relative weak phase is typically between tree level
and penguin amplitudes. Extensions to the Standard Model introduce additional amplitudes
with weak phases that can contribute to CP violation.
In the Standard Model, CP violation in the charm sector is small and $D^0\!-\!\overline D^0$ mixing is highly suppressed, so at current experimental sensitivities searches for CP violation in charm is for physics beyond the Standard Model.
%
Most CP violation results are from the FNAL fixed target experiments E791\cite{e791} and FOCUS\cite{focus}, and 
the CLEO\cite{cleo} experiment and search for direct CP violation. 
The CP violation asymmetry is defined as 
$A_{CP}\equiv \frac{\Gamma(D\to f) -\Gamma(\overline D \to \overline f)}{\Gamma(D\to f) +\Gamma(\overline D \to \overline f)}.
$
A few results from CLEO, BABAR\cite{babar} and BELLE\cite{belle} experiments consider CP violation in mixing.
\begin{figure}[t]
\vspace{6.5cm}
\includegraphics{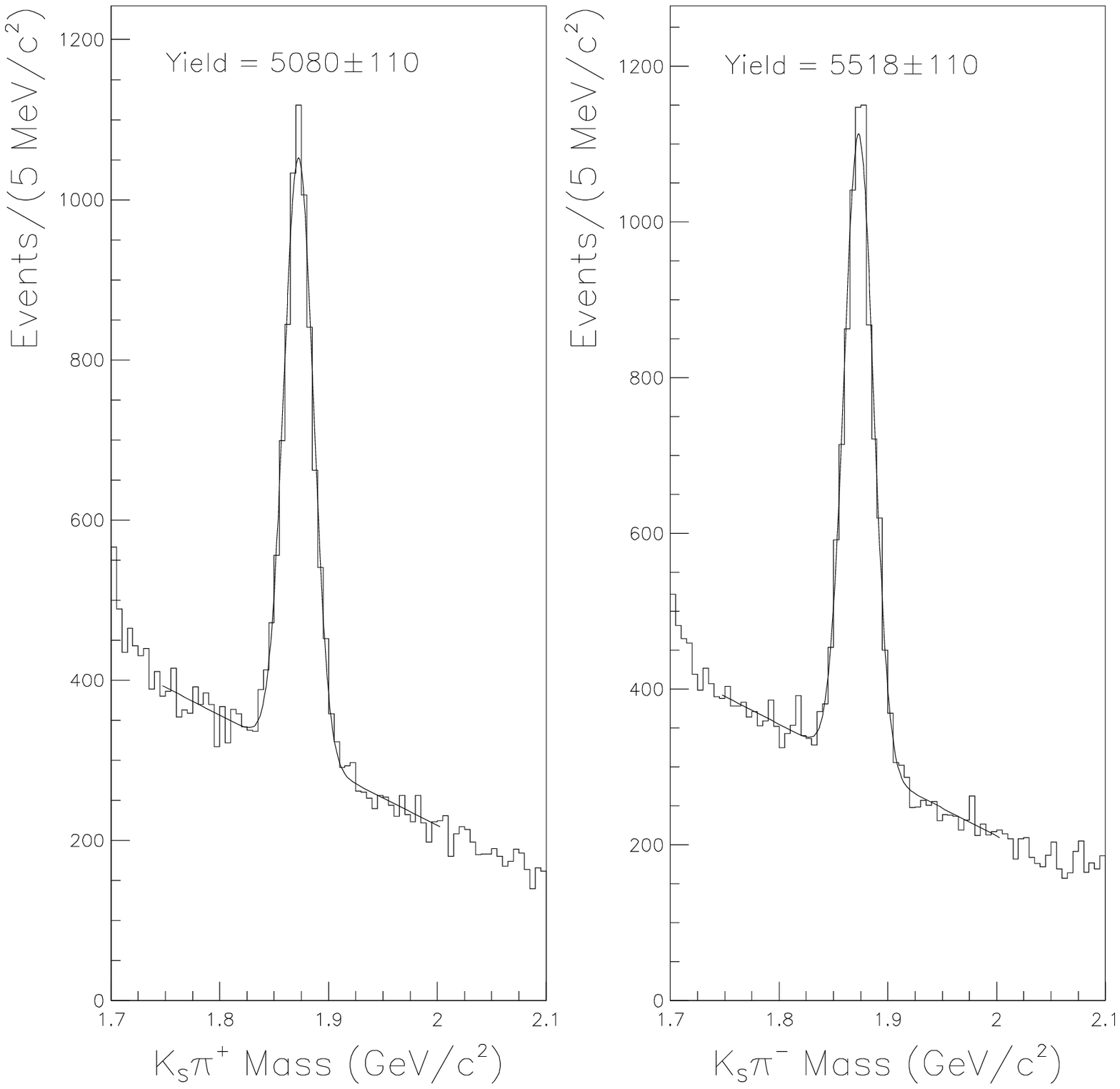}
\includegraphics{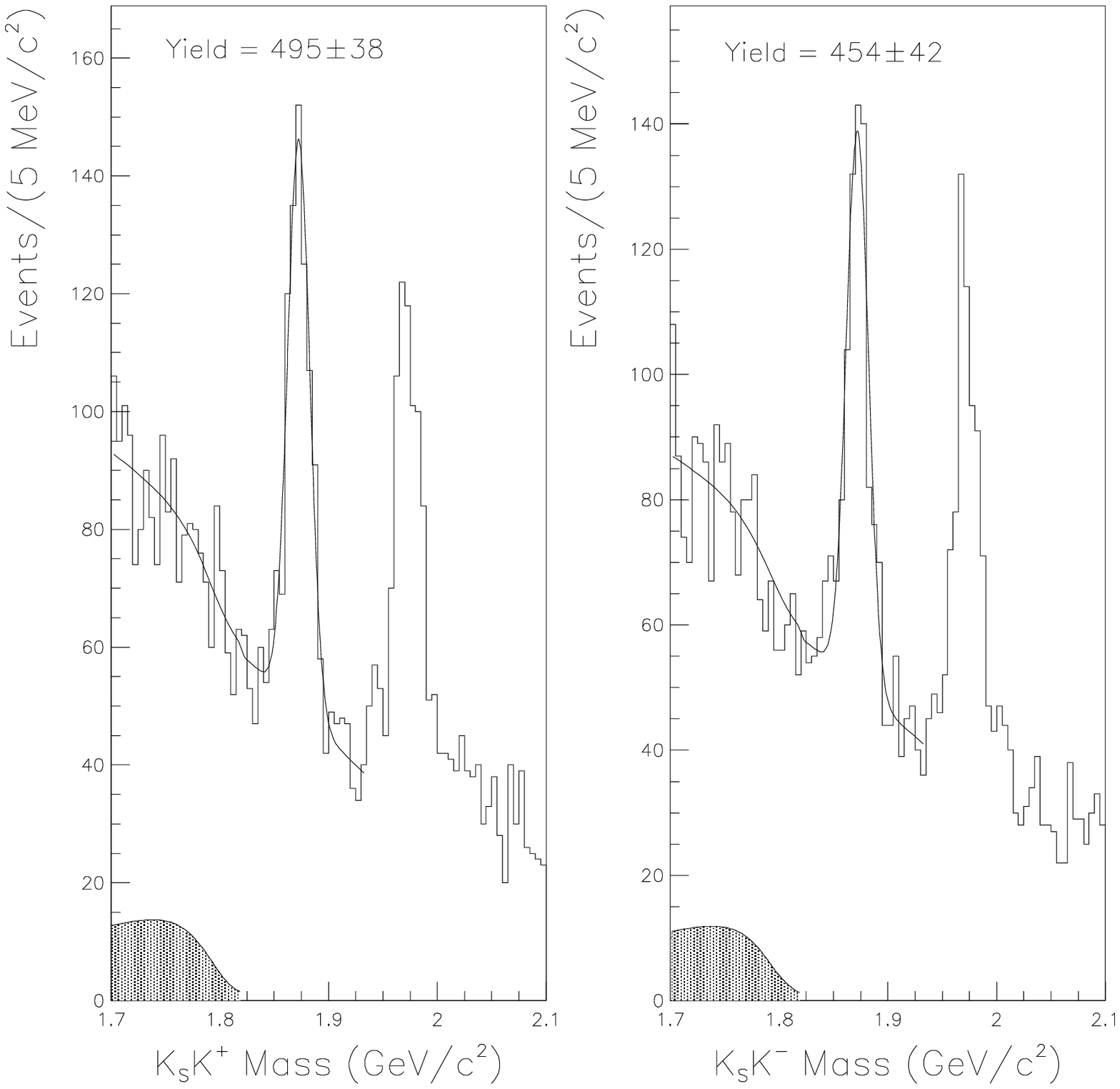}
 \caption{\it
      $D^+ \to K^0_S\pi^+, K^0_SK^+$ Mass plots.
    \label{fig:kspi} }
\end{figure}
\begin{table}[!]
\centering
\caption{ \it Branching Ratios (BR) and $A_{CP}$ of $D^+ \to K^0_S \pi^+,K^0_S K^+$.
}
\vskip 0.1 in
\begin{tabular}{lccc} \hline
 & FOCUS BR\cite{Link:2001zj} & PDG Average BR & $A_{CP}$\cite{Link:2001zj} \\ \hline
$\frac{\Gamma(\overline{K^0} \pi^+)}{\Gamma(K^- \pi^+ \pi^+)}$ & {$(30.60 \!\pm\! 0.46 \!\pm\!0.58)$\%$\!$} & $(32.0 \!\pm\! 4.0)$\% &  {$(-1.6 \!\pm\! 1.5 \!\pm\!0.9)$\%$\!\!$}\\
$\frac{\Gamma(\overline{K^0} K^+)}{\Gamma(K^- \pi^+ \pi^+)}$ & {$(6.04 \!\pm\! 0.35 \!\pm\!0.35)$\%$\!$} & $(7.7 \!\pm\! 2.2)$\%  & {$(6.9 \!\pm\! 6.0 \!\pm\!1.8)$\%$\!\!$}\\
$\frac{\Gamma(\overline{K^0} K^+)}{\Gamma(\overline{K^0} \pi^+)}$ & {$(19.96 \!\pm\! 1.20 \!\pm\!1.06)$\%$\!$} & $(26.3 \!\pm\! 3.5)$\%  & {$(7.1 \!\pm\! 6.1 \!\pm\!1.4)$\%$\!\!$}\\ 
\hline
\end{tabular}
\label{tab:ksh}
\end{table}


\section{Direct CP Violation}
\subsection{Two-body decays}
FOCUS has published results\cite{Link:2001zj} 
using the two-body decay modes $D^+\to K^0_S\pi^+$, where 
Cabibbo favored and doubly-Cabibbo suppressed amplitudes can interfere, and 
$D^+\to K^0_SK^+$ which is singly Cabibbo suppressed where interference between tree 
and penguin may occur. The
production mechanism in fixed target experiments yields different number of $D$ and 
$\overline D$ and so must normalize relative to another copious decay mode which is
unlikely to exhibit CP violation, in this case $D^+\to K^-\pi^+\pi^+$.
The $D^\pm \to K^0_S\pi^\pm, K^0_S\pi^\pm$ mass
plots are shown in Fig.~\ref{fig:kspi}. These decay modes will also manifest CP violation in $K^0\!-\!\overline K^0$ mixing.
The results tabulated in 
Table~\ref{tab:ksh} show no evidence for CP violation. This is consistent with Standard Model 
expectations $O(\sim 10^{-3})$.
%
%
\begin{table}[b]
\centering
\caption{ \it CP Asymmetry in Three-body Decays.}
\vskip 0.1 in
\begin{tabular}{lcc} \hline
 & {E791}\cite{Aitala:1996sh} & {FOCUS}\cite{Malvezzi:2002xt} \\ \hline
$A_{CP}(K^-K^+\pi^+)$ & $(-1.4 \pm 2.9)\%$ & $(0.6 \pm 1.1 \pm 0.5)\%$ \\
$A_{CP}(\phi\pi^+)$ & $(-2.8 \pm 3.6)\%$ & {Dalitz-plot analyses} \\
$A_{CP}(K^*K^+)$ & $(-1.0 \pm 5.0)\%$ & {in} \\
$A_{CP}(\pi^+\pi^-\pi^+)$ & $(-1.7 \pm 4.2)\%$ & {progress} \\
\hline
\end{tabular}
\label{tab:kkp}
\end{table}

\subsection{Three-body decays}
\begin{figure}[t]
\vspace{5.5cm}
\includegraphics{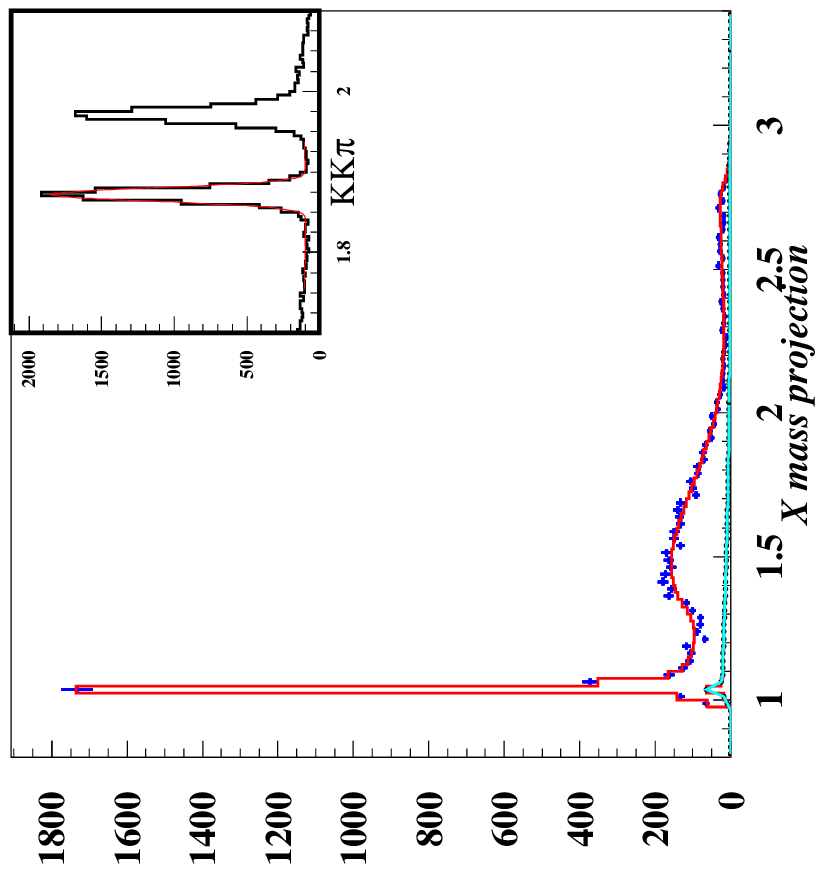}
\includegraphics{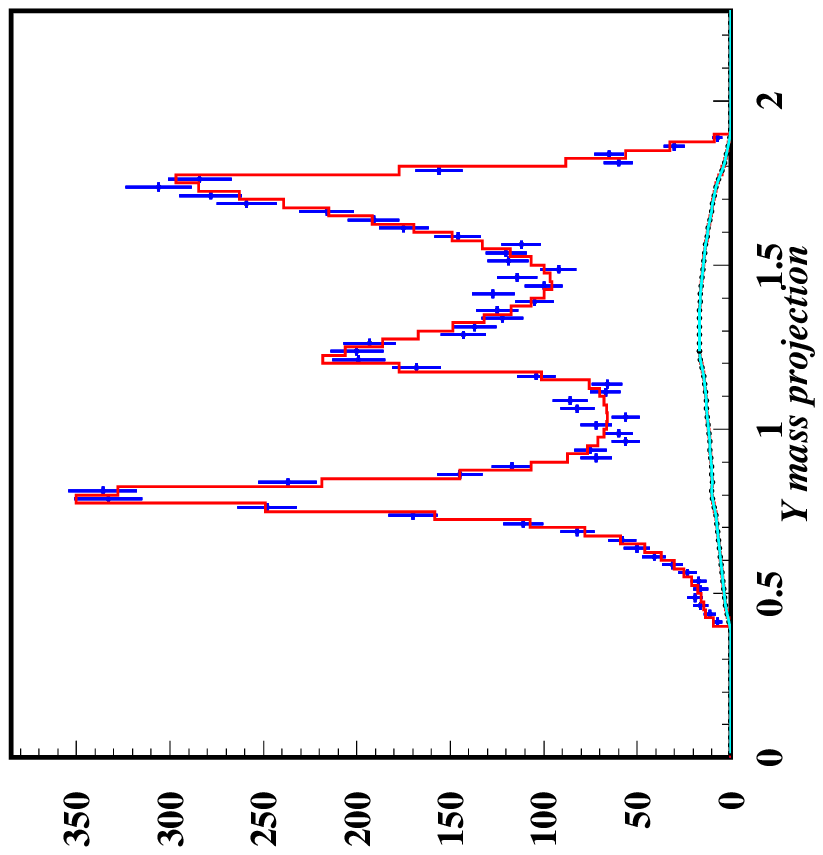}
\caption{\it FOCUS Dalitz-plot Analysis of $D^+\to K^+K^-\pi^+$\cite{Malvezzi:2002xt}: 
Projection of data (points) and fit (contour) for Left: $m^2_{KK}$ and Right: $m^2_{K\pi}$.}
\label{fig:focuskkp} 
\end{figure}

\begin{figure}[t]
\vspace{7.0cm}
\includegraphics{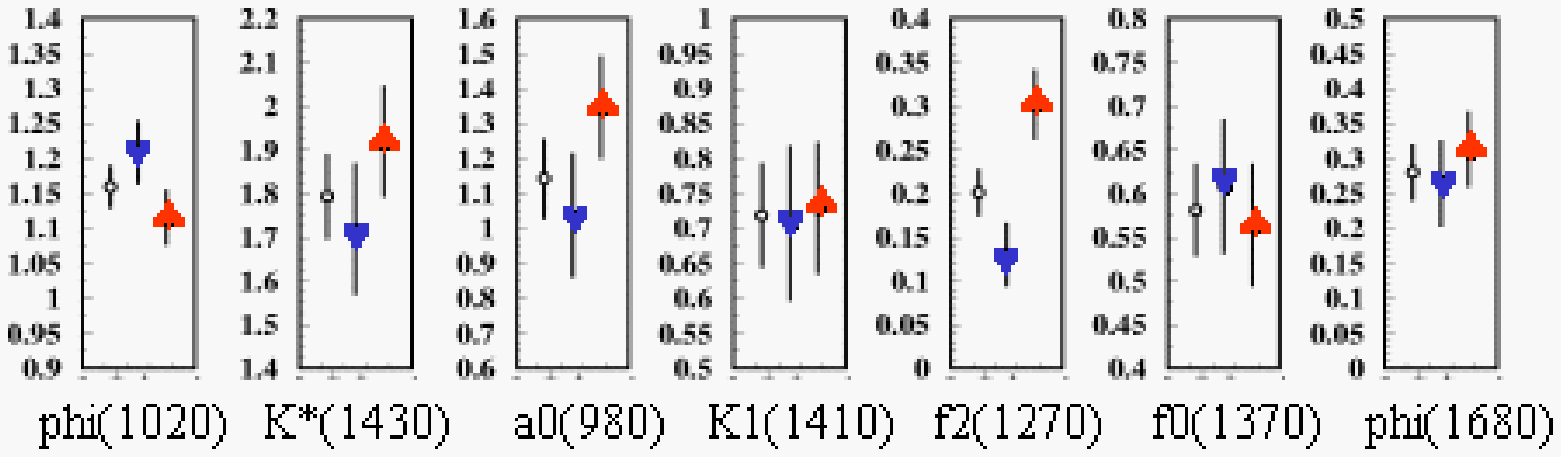}
\includegraphics{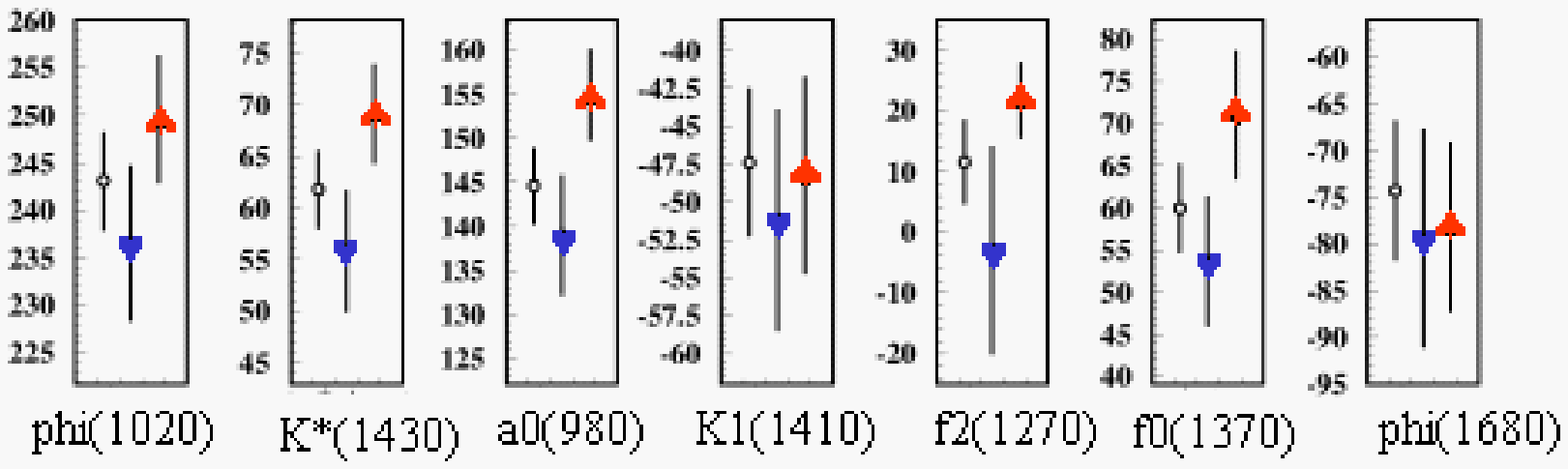}
\caption{\it FOCUS results for $D^+ \to K^+K^-\pi^+$\cite{Malvezzi:2002xt}. Amplitudes(top) and phases(bottom) of resonant substructure
for $D^\pm$(left), $D^+$(center), $D^-$(right).}
\label{fig:focuskkpcpv} 
\end{figure}

Direct CP violation searches in analyses of charm decays to three-body final states are
more complicated than two-body decays. Three methods have been used to search
for CP asymmetries. (1) Integrate over phase space and construct $A_{CP}$ as in two-body decays; (2) Examine CP asymmetry in the quasi-two-body resonances; (3) Perform a full Dalitz-plot 
analysis for $D$ and $\overline D$ separately. 
The Dalitz-plot analysis procedure~\cite{cleokpipi02,cleokspipi} 
allows increased sensitivity to CP violation by probing
decay amplitudes rather than the decay rate. Both E791\cite{Aitala:1996sh} and FOCUS have analyzed $D^+\to K^+K^-\pi^+$ using method (1). E791 has also analyzed $D^+\to K^-K^+\pi^+$ using method (2). These
results are shown in Table~\ref{tab:kkp}. 
FOCUS has a Dalitz-plot analysis in progress\cite{Malvezzi:2002xt}. The $D^+\to K^+K^-\pi^+$ 
Dalitz plot is well described by eight quasi-two-body decay channels. 
The projections of the data and fit are shown in Fig.~\ref{fig:focuskkp}.
A signature of CP violation in 
charm Dalitz-plot analyses is different amplitudes and phases for $D$ and $\overline D$ 
samples. The amplitudes and phases for
$D^+\to K^+K^-\pi^+$, $D^-\to K^-K^+\pi^-$ 
and the combined sample are shown graphically in Fig.~\ref{fig:focuskkpcpv}.
No evidence for CP violation is observed.


The decay $D^{*+} \to D^0\pi^+$ enables the discrimination between $D^0$ and ${\overline D}^0$.
The CLEO collaboration has searched for CP violation integrated across the Dalitz plot
in $D^0 \to K^\mp\pi^\pm\pi^0$, $K^0_S\pi^+\pi^-$ and $\pi^+\pi^-\pi^0$ decays.
The integrated CP violation across the Dalitz plot is determined by
\begin{equation}
{\cal A}_{CP} = \int {\frac{\left|{\cal M}_{D^0}\right|^2 - \left|{\cal M}_{\overline D^0}\right|^2}{\left|{\cal M}_{D^0}\right|^2 + \left|{\cal M}_{{\overline D}^0}\right|^2}dm^2_{ab}dm^2_{bc}} / \int dm^2_{ab}dm^2_{bc}.
\end{equation}
The CLEO results for integrated CP asymmetry in $D^0$ decays
are reported in Table~\ref{tab:cleocpv}. No evidence of CP violation has been observed.
\begin{table}[b]
\centering
\caption{ \it Integrated CP Asymmetry in Dalitz-plot Analysis.}
\vskip 0.1 in
\begin{tabular}{lcc} \hline
& Decay Mode & ${\cal A}_{CP}$(\%)  \\ \hline
{CLEO}\cite{cleokpipi02} & $D^0 \to K^- \pi^+\pi^0$ & $-3.1\pm 8.6$ \\
{CLEO}\cite{cleokpipi0} & $D^0 \to K^+ \pi^-\pi^0$ & $9^{+22}_{-25}$ \\
{CLEO}\cite{cleokspipi} & $D^0 \to K^0_S \pi^+\pi^-$ & $\!-0.9\! \pm\! 2.1 ^{+1.0}_{-4.3}\!^{+1.3}_{-3.7}$ \\
{CLEO}\cite{cleopipipi0} & $D^0 \to \pi^+\pi^-\pi^0$ & $1^{+9}_{-7}\pm 9$ \\ \hline
\end{tabular}
\label{tab:cleocpv}
\end{table}

\begin{figure}[t]
\vspace{5.75cm}
\includegraphics{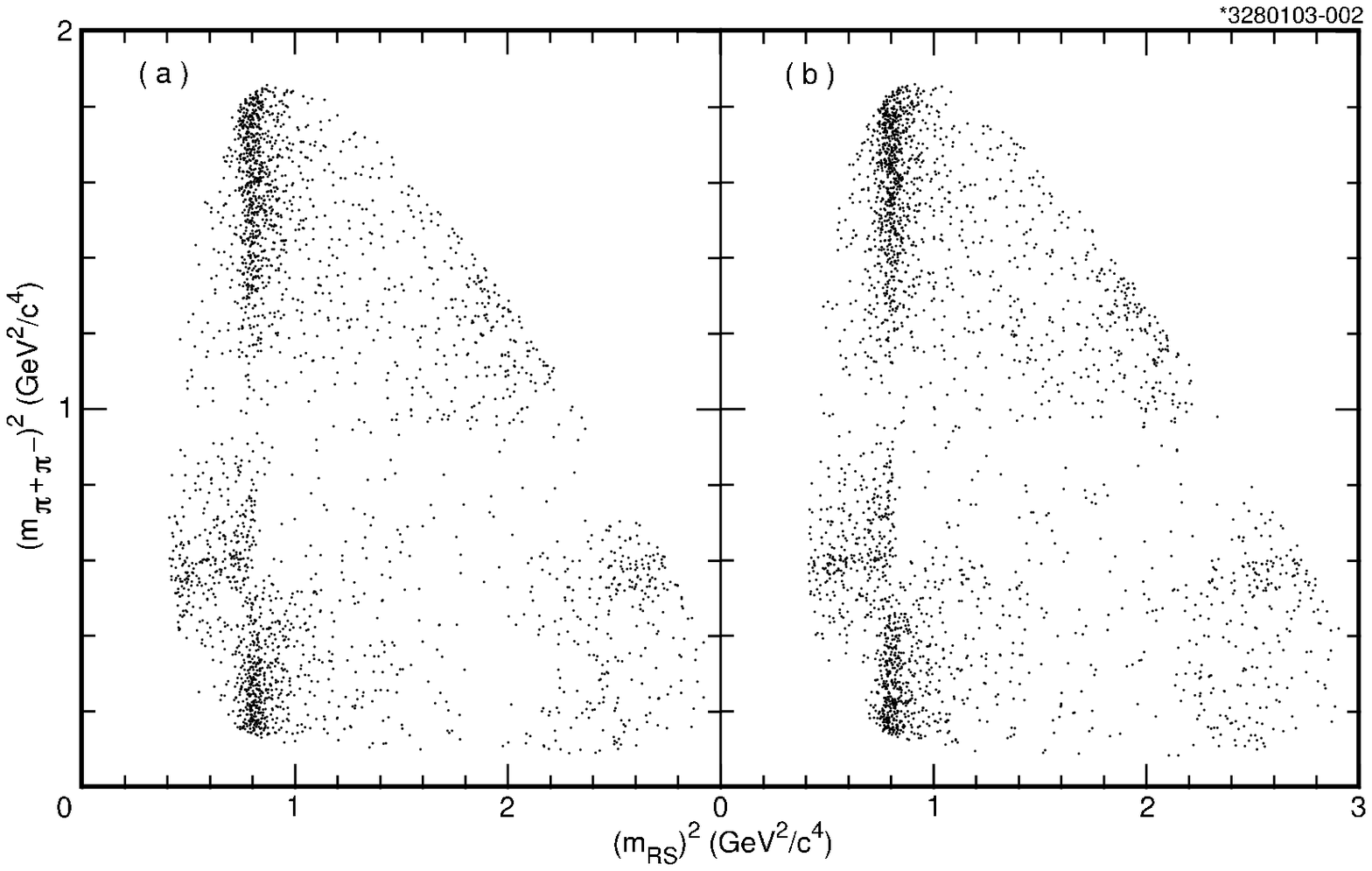}
 \caption{\it CLEO~II.V: $D^0 \to K^0_S\pi^+\pi^-$ and $\overline D^0 \to K^0_S\pi^+\pi^-$ Dalitz plots\cite{cleokspipi}.
    \label{fig:cleokspipi} }
\end{figure}

CLEO has considered CP violation more generally in a simultaneous fit to the $D^0\to K^0_S\pi^+\pi^-$ and $\overline D^0 \to K^0_S\pi^+\pi^-$ Dalitz plots, shown in Fig.~\ref{fig:cleokspipi}.
In the isobar model\cite{cleokpipi02}, each resonance, $j$, 
has its own amplitude, $a_j$, and phase, $\delta_j$. 
A second process, not necessarily of Standard Model origin, 
is allowed to contribute to each $j$-th resonance. 
In general, the amplitudes to the $j$-th quasi-two-body state can be expressed as
\begin{equation}
(a_j e^{i(\delta_j\pm\phi_j)}\!\pm\! b_j e^{i(\delta_j\pm\phi_j)}){\cal A}_j\! =\! a_j e^{i(\delta_j\pm\phi_j)}(1\! \pm\! \frac{b_j}{a_j}){\cal A}_j,
\label{eqn:cpvpar}
\end{equation}
with `$+$' for $D^0$ and `$-$' for $\overline D^0$ and
${\cal A}_j = {\cal A}_j(m_{K^0_S\pi}^2,m_{\pi\pi}^2)$ is the spin-dependent Breit-Wigner 
amplitude for resonance $j$ as
described in Ref.~\cite{cleokpipi02}. Thus $a_j$ and
$\delta_j$ are explicitly CP conserving amplitude and phase,
$b_j$ is an explicity CP violating amplitude normalized by the CP
conserving amplitude $a_j$, and $\phi_j$ is an explicitly CP
violating phase. In the absence
of CP violation $b_j$ and $\phi_j$ would be zero. 
The results of the fit to the $D^0$ and $\overline D^0\to K^0_S\pi^+\pi^-$ Dalitz plots 
are consistent with each other and with no CP violation.
The fractional CP violating amplitude and CP violating phase,
$b_j/a_j$ and $\phi_j$ are given in Table~\ref{tab:cleokspipi}.

\begin{table}[!]
\centering
\caption{ \it CLEO~II.V: CP Asymmetry in $D^0 \to K^0_s\pi^+\pi^-$\cite{cleokspipi}.}
\vskip 0.1 in
\begin{tabular}{lcc}
Component     & Amplitude Ratio\,($b_j/a_j$)& 
Phase\,($\phi_j$)\\ \hline
$K^\ast(892)^+\pi^{\rm -}, K^\ast(892)^+\!\to\! K^0 \pi^+$ & $-0.12\!^{+0.21}_{-0.22}\!^{+0.09}_{-0.15}\!^{+0.11}_{-0.03}$&$6\!^{+ 21}_{- 22}\!^{+ 13}_{- 35}\!^{+ 18}_{-  4}$ \\
$\overline{K}^0 \rho^0$                                                         &$0.001\!\pm\!0.022\!^{+0.011}_{-0.009}\!^{+0.002}_{-0.011}$&-$1\!^{+ 16}_{- 18}\!^{+  9}_{- 31}\!^{+ 21}_{-  3}$ \\
$\overline{K}^0 \omega, \omega \!\to\! \pi^+\pi^{\rm -}$                         &-$0.14\!^{+0.10}_{-0.11}\!^{+0.11}_{-0.01}\!^{+0.01}_{-0.02}$&-$8\!^{+ 17}_{- 19}\!^{+  8}_{- 30}\!^{+ 20}_{-  3}$ \\
$K^\ast(892)^{\rm -}\pi^+, K^\ast(892)^{\rm -}\!\to\!{\overline K}^0\pi^{\rm -}$             &-$0.002\!\pm\!0.012\!^{+0.008}_{-0.003}\!^{+0.002}_{-0.002}$&-$3\!^{+ 16}_{- 18}\!^{+  9}_{- 30}\!^{+ 21}_{-  3}$ \\
$\overline{K}^0 f_0(980), f_0(980) \!\to\! \pi^+ \pi^{\rm -}$                    &-$0.04\!\pm\!0.06\!^{+0.13}_{-0.04}\!^{+0.00}_{-0.04}$&$9\!^{+ 16}_{- 17}\!^{+ 10}_{- 29}\!^{+ 20}_{-  3}$ \\
$\overline{K}^0 f_2(1270), f_2(1270) \!\to\! \pi^+ \pi^{\rm -}$                  &$0.16\!^{+0.28}_{-0.27}\!^{+0.15}_{-0.37}\!^{+0.11}_{-0.18}$&$22\!^{+ 19}_{- 20}\!^{+ 12}_{- 32}\!^{+ 20}_{-  2}$\\
$\overline{K}^0 f_0(1370), f_0(1370) \!\to\! \pi^+ \pi^{\rm -}$                  &$0.08\!^{+0.06}_{-0.05}\!^{+0.01}_{-0.11}\!^{+0.06}_{-0.03}$&$8\!^{+ 15}_{- 17}\!^{+ 10}_{- 28}\!^{+ 20}_{-  4}$\\
$K^\ast_0(1430)^{\rm -}\pi^+, K^\ast_0(1430)^{\rm -}\!\to\!\overline{K}^0\pi^{\rm -}$        &-$0.02\!\pm\!0.06\!^{+0.04}_{-0.02}\!^{+0.00}_{-0.01}$&-$3\!^{+ 17}_{- 19}\!^{+ 13}_{- 36}\!^{+ 23}_{-  2}$\\
$K^\ast_2(1430)^{\rm -}\pi^+, K^\ast_2(1430)^{\rm -}\!\to\!\overline{K}^0\pi^{\rm -}$        &-$0.05\!\pm\!0.12\!^{+0.04}_{-0.14}\!^{+0.04}_{-0.00}$&$3\!^{+ 17}_{- 18}\!^{+ 10}_{- 31}\!^{+ 21}_{-  2}$\\
$K^\ast(1680)^{\rm -} \pi^+, K^\ast(1680)^{\rm -} \!\to\! \overline{K}^0 \pi^{\rm -}$        &-$0.20\!^{+0.28}_{-0.27}\!^{+0.05}_{-0.22}\!^{+0.02}_{-0.01}$&-$3\!^{+ 19}_{- 20}\!^{+ 20}_{- 25}\!^{+ 27}_{-  2}$\\ \hline
\end{tabular} 
\label{tab:cleokspipi}
\end{table}
\subsection{Four-body decays}
\begin{figure}[!]
\vspace{6.cm}
\includegraphics{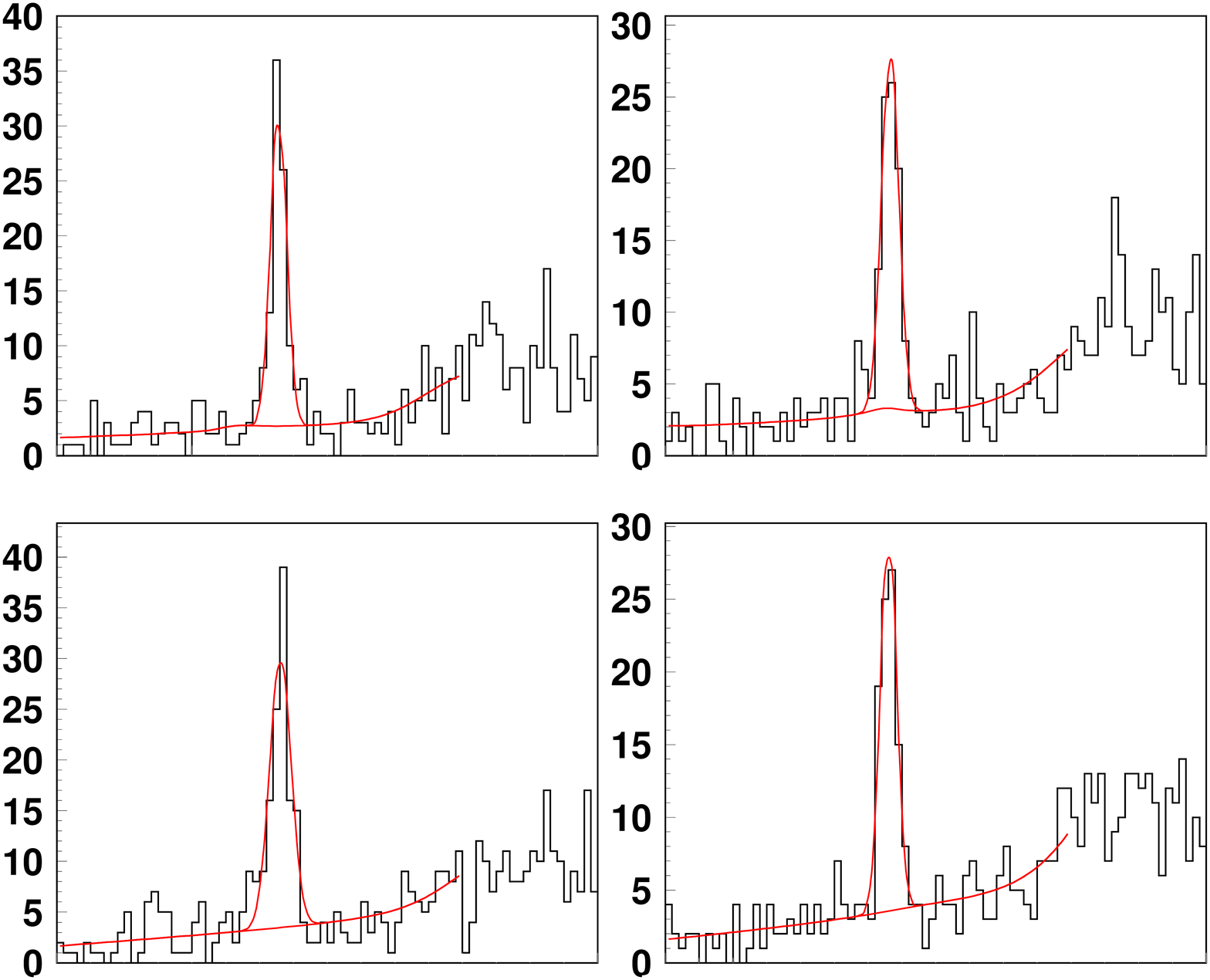}
 \caption{\it
Top (Bottom): $D^0$($\overline D^0$) $m_{K^-K^+\pi^-\pi^+}$ for 
Left (Right): $C_T<0$($>0$).}
\label{fig:Tviol} 
\end{figure}

FOCUS has searched for T-violation using the four-body decay modes $D^0\to K^+K^-\pi^+\pi^-$
\cite{focustv}.
A T-odd correlation can be formed wit the momenta, $C_T\equiv (\vec{p}_{K^+}.(\vec{p}_{\pi^+}\times \vec{p}_{\pi^-}))$. Under time-reversal, $C_T \to -C_T$, however $C_T\ne 0$ does not
establish T-violation. Since time reversal is implemented by an anti-unitary operator, $C_T\ne 0$, can be induced by FSI\cite{bigisanda}. This ambiguity can be resolved by measuring 
$\overline C_T \equiv (\vec{p}_{K^+}.(\vec{p}_{\pi^+}\times \vec{p}_{\pi^-}))$ in 
$\overline D^0\to K^+K^-\pi^+\pi^-$; $C_T\ne \overline C_T$ establishes T violation. FOCUS
reports a preliminary asymmetry $A_T = 0.075\pm 0.064$ from a sample of $\sim 400$ decays. 
The mass distributions for $D^0$ and $\overline D^0$ for $C_T$ and $\overline C_T$ greater than
and less than zero are shown in Fig.~\ref{fig:Tviol}.

\section{CP Violation in $D^0\!-\!\overline D^0$ Mixing}
E791\cite{Aitala:1997ff}, FOCUS\cite{Link:2000aw} and
CLEO\cite{cleokkpipi} 
have all searched for CP violation in the Cabibbo suppressed decays
to CP eigenstates,
$D^0 \to K^+K^-$ and $D^0\to \pi^+\pi^-$.
These measurements, tabulated in 
Table~\ref{tab:2bodyD0}, are approaching the 1\% level, where non-Standard Model physics
may appear. 
\begin{table}[t]
\centering
\caption{ \it CP Asymmetry in $D^0 \to K^+K^-, \pi^+\pi^-$.}
\vskip 0.1 in
\begin{tabular}{lcc} \hline
Expt & $A_{CP}(KK)$ (\%) & $A_{CP}(\pi\pi)$ (\%) \\ \hline
{E791}\cite{Aitala:1997ff}   & $-1.0 \pm 4.9 \pm 1.2$ & $-4.9 \pm 7.8 \pm 3.0$ \\
{FOCUS}\cite{Link:2000aw} & $-0.1 \pm 2.2 \pm 1.5$ & $\;\;\;4.8 \pm 3.9 \pm 2.5$ \\
{CLEO}\cite{cleokkpipi} & $\:\;\;0.0 \pm 2.2 \pm 0.8$ & $\;\;\;1.9 \pm 3.2 \pm 0.8$ \\ \hline
Expt & Mode(s) & $\Im(x)$ (\%)  \\ \hline
{BELLE}\cite{bellekk}  &$K^+K^-$  & $-0.20 \pm 0.63 \pm 0.30$ \\
{BABAR}\cite{babarkkpipi} & $K^+K^-,\pi^+\pi^-$ & $-0.8 \pm 0.6 \pm 0.2$  \\ \hline
\end{tabular}
\label{tab:2bodyD0}
\end{table}

Time dependent $A_{CP}$ measurements performed by BABAR\cite{babarkkpipi} and
BELLE\cite{bellekk} 
can distinguish
direct and indirect CP violation. Since mixing is small the decay time to CP eigenstates can
be fit with a single exponential $\exp{[-\Gamma (1 + y \mp \Im(x))]}$. The signature of CP 
violation is $D^0$ and $\overline D^0$ having different decay rates, $\Im(x)\ne 0$, to CP 
eigenstates. The results are tabulated in 
Table~\ref{tab:2bodyD0} and are consistent with the absence of CP violation.




\section{Summary and Future Outlook}
Searches for CP violation in charm decay at fixed target and $e^+e^-$ facilities are
summarized in Table~\ref{tab:fixedtargetcpv} and \ref{tab:epemcpv}, respectively, including
additional results not discussed in the text. 
FOCUS and CLEO continue work on studying CP violation using Dalitz-plot analyses, $D^+\!\to\!K^+K^-\pi^+, \pi^+\pi^-\pi^+$ and $D^0\!\to\!K^0_S\pi^0\pi^0$, respectively. BABAR and BELLE have
each accumulated twenty-five times the statistics of CLEO~II.V, approaching sensitivity to
CP violation in Kaon mixing, in modes like $D \to K^0_S\pi$. Presently
CLEO-c\cite{cleoc} is taking data at the $\psi(3770)$ with the goal of accumulating 18 million $D\overline D$ events and attain sensitivity 
comparable to 1~ab$^{-1}$ of B-factory data. In addition, 
CLEO-c will exploit the CP coherent $D\overline D$ system to probe CP violation.
Beginning in 2009 the BTeV experiment\cite{BTeV} will start to accumulate $\sim\!1000\times$ the 
charm statistics of FOCUS opening up a new regime in charm CP and T violation studies.
\begin{table}[t]
\centering
\caption{ \it Fixed Target Experiments: CP Violation Searches in Charm.}
\vskip 0.1 in
\begin{tabular}{l|cc}
$A_{CP}$ mode & {E791}(\%) & {FOCUS}(\%) \\ \hline
$\!D^0 \!\rightarrow\! K^-\!K^+$ & $-1.0 \!\pm\! 4.9 \!\pm\! 1.2$\cite{Aitala:1997ff} & $-0.1 \!\pm\! 2.2 \!\pm\! 1.5$\cite{Link:2000aw} \\
$\!D^0 \!\rightarrow\! \pi^-\!\pi^+$ & $-4.9 \!\pm\! 7.8 \!\pm\! 3.0$\cite{Aitala:1997ff} & $4.8 \!\pm\! 3.9 \!\pm\! 2.5$\cite{Link:2000aw} \\
$\!D^+ \!\!\rightarrow\! K^0_{\!S}\pi^+$ & & $-1.6 \!\pm\! 1.5 \!\pm\! 0.9$\cite{Link:2001zj} \\
$\!D^+ \!\!\rightarrow\! K^0_{\!S}K^+$ & & $6.9 \!\pm\! 6.0 \!\pm\! 1.8$\cite{Link:2001zj}\\
$\!D^+ \!\!\rightarrow\! K^-\!K^+\!\pi^+\!$ & $-1.4 \!\pm\! 2.9$\cite{Aitala:1996sh} & $0.6 \!\pm\! 1.1 \!\pm\! 0.5$\cite{Malvezzi:2002xt} \\
$\!D^+ \!\!\rightarrow\! \phi\pi^+$ & $-2.8 \!\pm\! 3.6$\cite{Aitala:1996sh} & \\
$\!D^+ \!\!\rightarrow\! K^*K^+$ & $-1.0 \!\pm\! 5.0$\cite{Aitala:1996sh} & \\
$\!D^+ \!\!\rightarrow\! \pi^-\!\pi^+\pi^+$ & $-1.7 \!\pm\! 4.2$\cite{Aitala:1996sh} & \\
\end{tabular}
\label{tab:fixedtargetcpv}
\end{table}
\begin{table}[t]
\centering
\caption{ \it $e^+e^-$ Experiments: CP Violation Searches in Charm.}
\vskip 0.1 in
\begin{tabular}{l|ccc}
$A_{CP}$ mode & {CLEO}
 & {BABAR}(\%) & {BELLE}(\%) \\ \hline
$\!D^0 \!\rightarrow\! K^+\!\pi^-$  & $2^{+19}_{-20}$\cite{cleokpi} & $9.5\pm 10.3$\cite{babarkpi}  & \\
$\!D^0 \!\rightarrow\! K^+\!\pi^-\pi^0$  &  $9^{+25}_{-22}$\cite{cleokpipi0} &  & \\
$\!D^0 \!\rightarrow\! K^-\!K^+$ & $0.0 \!\pm\! 2.2 \!\pm\! 0.8$\cite{cleokkpipi} &  ${-0.8\pm0.6}$\cite{babarkkpipi} &  $0.2 \!\pm\!0.7$\cite{bellekk}\\ 
$\!D^0 \!\rightarrow\! \pi^-\!\pi^+$ & $1.9 \!\pm\! 3.2 \!\pm\! 0.8$\cite{cleokkpipi} &  {${-0.8\pm0.6}$}\cite{babarkkpipi}  & \\
$\!D^0 \!\rightarrow\! \pi^0\pi^0$ & $0.1 \!\pm\! 4.8$\cite{cleoDcpv} & & \\
$\!D^0 \!\rightarrow\! K^0_{\!S}K^0_{\!S}$ & $-23 \!\pm\! 19$\cite{cleoDcpv} & & \\
$\!D^0 \!\rightarrow\! K^0_{\!S}\pi^0$ & $0.1 \!\pm\! 1.3$\cite{cleoDcpv} & & \\
$\!D^0 \!\rightarrow\! K^0_{\!S}\pi^+\pi^-$ & $-3.9^{+4.6}_{-4.9}$\cite{cleokspipi} & & \\
$\!D^0 \!\rightarrow\! K^0_{\!S}\phi$ & $2.8 \!\pm\! 9.4$\cite{cleoksphi} & & \\
$\!D^0 \!\rightarrow\! K^-\!\pi^+\pi^0$ & $-3.1 \!\pm\! 8.6$\cite{cleokpipi02} & & \\
$\!D^0 \!\!\rightarrow\! \pi^+\!\pi^-\pi^0$ & $-1^{+13}_{-11}$\cite{cleopipipi0} & & \\ \hline
\end{tabular}
\label{tab:epemcpv}
\end{table}

\end{document}